# Investigation of Defect Modes of Chiral Photonic Crystals

M.A. Kutlan

Institute for Particle & Nuclear Physics, Budapest, Hungary

kutlanma@gmail.com

**Abstract**. Some properties of defect modes of cholesteric liquid crystals (CLC) are presented. It is shown that when the CLC layer is thin the density of states and emission intensity are maximum for the defect mode, whereas when the CLC layer is thick, these peaks are observed at the edges of the photonic band gap. Similarly, when the gain is low, the density of states and emission intensity are maximum for the defect mode, whereas at high gains these peaks are also observed at the edges of the photonic band gap. The possibilities of low-threshold lasing and obtaining high-$Q$ microcavities have been investigated.

1. INTRODUCTION

Liquid crystals are an interdisciplinary research area. Chemists are creating new materials and they are using the supramolecular system for synthesis, analysis and purifications of other materials. A special focus nowadays is the template synthesis of nanoscience materials. Biologist are studing the cell membrane as a liquid crystal with respect to intercellular transport and communications. Physicist are measuring the unique properties of liquid crystals, e.g. ferroelectricity, piezoelectricity, reentrante phases, frustrated phases, phase transitions etc., and they are developing new ways of electrooptics. Engineers are creating electrooptical displays (LCD, STN, TFT), optical information storage, switchable gearboxes, surfactants and lubricants etc. Liquid crystals cover a wide range of chemical structures, including rod-like molecules, disc-like molecules, amphiphilic materials, cellulose derivatives, metallomesogens, steroids, glycolipids, etc. Typical applications for liquid crystals are LCD displays, surfactants, membranes, color pigments, advanced materials, photoconductors, materials for lasers and etc.

Cholesteric liquid crystals (CLCs) are the most widespread representatives of 1D chiral photonic crystals (CPCs) due to the possibility of spontaneous selforganization of their periodic structure and controllability of their photonic band gap (PBG) in a wide frequency range. Intensive studies in the field of CLC field are under way. CLCs with defects of various types in their structure have been considered recently in view of generating additional resonant modes in CLCs and studying the possibility of low-threshold lasing on these modes [1-20]. CLCs with an isotropic defect were considered in [21-96] and references therein.

In this paper we investigated the features of the photonic density of states (PDOS), $Q$ factor, and emission from CLCs with an isotropic defect and analyzed the effect of the defect-layer thickness, the thickness of the system as a whole, the position of the defect layer in the system, and the dielectric boundaries on the specific features of the defect modes.

## 2. THE METHOD OF ANALYSIS AND RESULTS

A CLC with an isotropic defect can be considered as a three-layer system composed of two CLC layers (CLC(1) and CLC(2)) with an isotropic dielectric layer (IDL) between. Let us consider the normal incidence of light on the CLC(1)–IDL–CLC(2) multilayer system from the left side. Calculations were performed for a CLC layer with $n_o = \sqrt{\varepsilon_1} = 1.4639$ and $n_e = \sqrt{\varepsilon_2} = 1.5133$ (the CLC composition is cholesteryl nonanoate : cholesteryl chloride : cholesteryl acetate = 20 : 15 : 6), which has a helical pitch $p = 0.42$ μm in the optical range at room temperature (24°C). The CLC spiral is right-handed; therefore, there is a PBG for right-handed circularly polarized light incident on the defect-free CLC layer and no such a band for left-handed circularly polarized light.

Now we consider how the position of the defect layer in the system affects the defect mode features. Figure 1 shows the dependences of (a) the defect mode wavelengths $\lambda^d$, (b) the reflectances $R$, (c) the $Q$ factors, (d) the relative PDOS $\rho_m / \rho_{iso}$, and (e) the emission intensities $|A|$ (at the defect mode) on the defect position in the structure ($x/\sigma$, where $x$ is counted from the midplane of the CLC layer) for right handed (solid line) and left handed (dashed line) circularly polarized light incident on the system.

As can be seen in the figure, peaks of the reflectance $R$ and emission intensity for right-handed circularly polarized incident light are observed, not when the defect lies in the midplane of the system, but when it is slightly shifted to the left. A similar situation occurs in the case of anisotropic defect but for both circular polarizations [7]. The asymmetry in the dependence of the reflection on the defect position indicates the presence of nonreciprocity in the system under consideration. Since the nonreciprocity is rather high, one can use this system as an optical diode.

Concerning the $Q$ factors and relative PDOS $\rho_m / \rho_{iso}$, these parameters are extremum when the defect is in the midplane of the system.

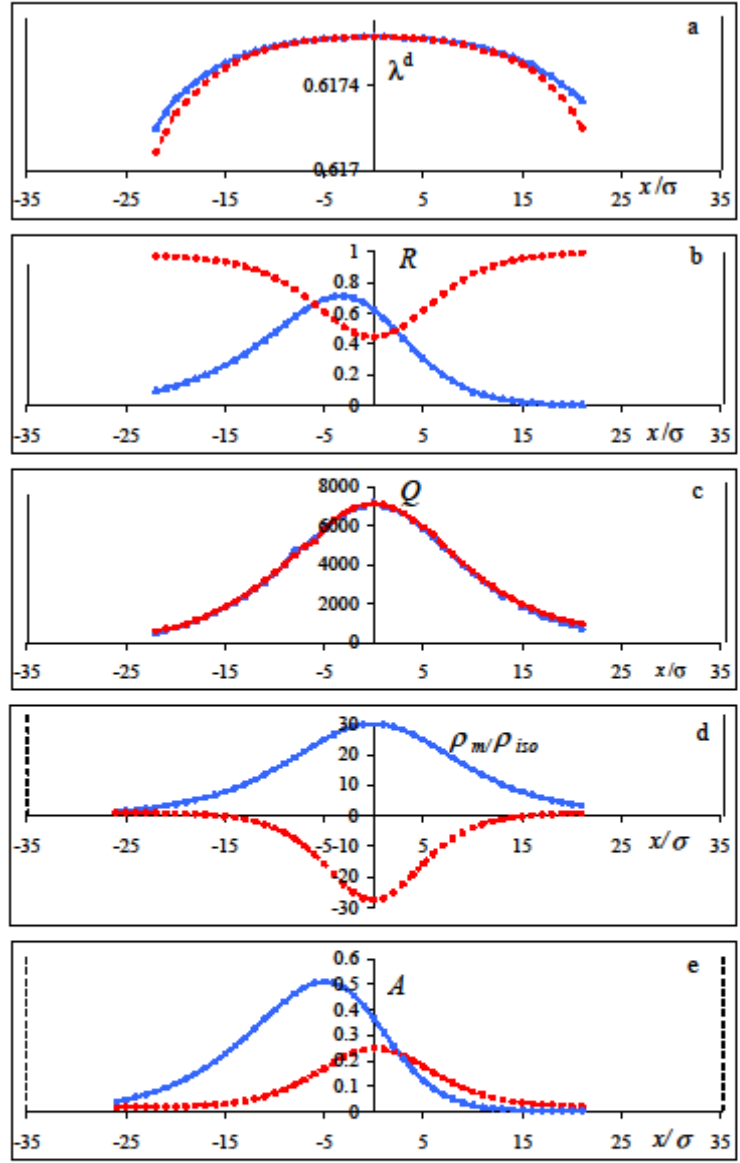

**Fig. 1.** Dependences of (a) defect-mode wavelengths $\lambda^d$ and (b) reflectances $R$, (c) $Q$ factors, (d) relative PDOSs ρm/ρiso, and (e) emission intensities $A$ at defect modes on defect position ($x/\sigma$) ($x$ is counted from the midplane of the system). Parameters of system and curves are are as follows: ε1 = 2.29–0.0001$i$, ε2 = 2.143–0.0001$i$, $Sd$ = 0, helical pitch σ = 0.42 μm, CLC thickness $d$ = 70σ, defect-layer thickness $dd$ = 1.86 μm, defect-layer refractive index $nd$ = 1.7, and refractive index of the medium $n0$ = $\sqrt{(\varepsilon_1 + \varepsilon_2)/2}$ around system..

Thus, the defect modes are highly excited when the defect layer is near the midplane of the system but barely manifest themselves when the defect is at the boundaries. The defect mode with left-handed circular polarization is maximally excited when the defect is in the midplane, whereas the defect mode with right-handed circular polarization is maximally excited when the defect is slightly shifted to the left.

Below we will discuss the effect of dielectric boundaries on the defect mode.

For clarity, Fig. 2a shows the 3D dependences of the transmittance $T$ on the wavelength $\lambda$ and $n0$, and Fig. 2b shows the dependence of the reflectance $R$ on the wavelength $\lambda$ and $n0$. The light incident on the system is either (a) right - handed or (b) left - handed circularly polarized. As can be seen in the figures, the Fresnel reflections significantly affect the reflection spectra, especially for left-handed circularly polarized light. A comparison of the reflection spectra at different values of the parameter α shows that the amplification of Fresnel reflections (i.e., the increase in the difference between α and unity) reduces the dip of the defect mode for incident right-handed circularly polarized light and increases the defect-mode peak for left-handed circularly polarized light.

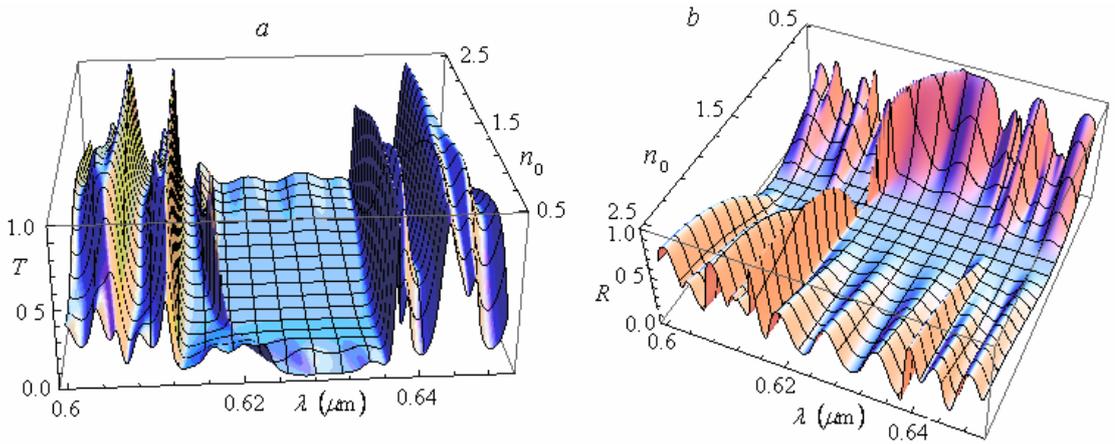

**Fig. 2.** 3D dependences of (a) transmittance $T$ and (b) reflectance $R$ on wavelength $\lambda$ and $n0$. Incident light is (a) right-handed and (b) left-handed circularly polarized $\varepsilon_1'' = \varepsilon_2'' = 0$, $n^d = \sqrt{(\varepsilon_1 + \varepsilon_2)/2}$, $d^d$ = 0.25 mm. Other parameters of system and curves are same as in Fig. 1.

As was shown in [15], the situation is the same for the increase in the CLC layer thickness. A change in $n0$ also changes the defect-mode wavelength. At $\Delta n = 0$, the defect mode with left-handed circular polarization is completely suppressed; i.e., it does not manifest itself. This is natural because the CLC with a defect in its structure is nothing more but a microcavity, and multiple reflections are necessary to induce resonant modes.

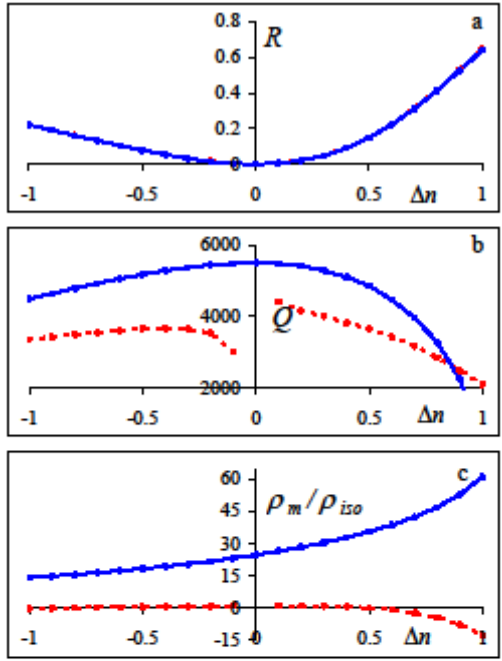

**Fig. 3.** Dependences of (a) reflectances $R$, (b) $Q$ factors, and (c) relative PDOSs $\rho_m/\rho_{iso}$ at defect modes on $\Delta n = \sqrt{(\varepsilon_1+\varepsilon_2)/2} - n_0$, $\varepsilon_1'' = \varepsilon_2'' = 0$, $n^d = \sqrt{(\varepsilon_1+\varepsilon_2)/2}$, $d^d = 0.25$ mm. Parameters of system and curves are same as in Fig. 1.

Figure 3 shows the dependences of (a) the reflectances $R$, (b) the $Q$ factors, and (c) the relative PDOSs $\rho_m/\rho_{iso}$ (at the defect modes) on $\Delta n$ for incident light with right-handed (solid line) and left-handed (dashed line) circular polarizations. Here $\Delta n = \sqrt{(\varepsilon_1+\varepsilon_2)/2} - n_0$, (ε1 and ε2 are the principal values of the CLC dielectric tensor, and $n0$ is the refractive index of the medium around the system. In addition, we assumed the refractive index of the isotropic defect to be $n^d = \sqrt{(\varepsilon_1+\varepsilon_2)/2}$.

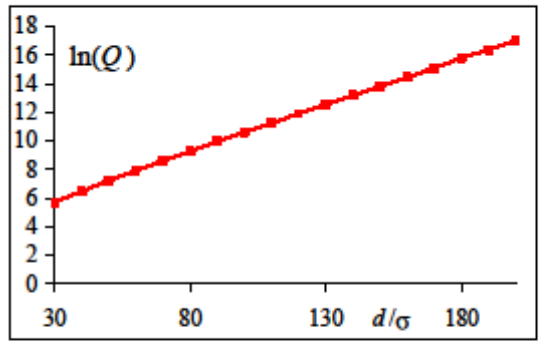

**Fig. 4.** Dependence of ln$Q$ on CLC layer thickness ($d/\sigma$). Parameters of system are same as in Fig. 1.

It follows from these results that the $Q$ factor for right_handed circularly polarized light is maximum at $\Delta n = 0$.

Since microcavities with a high $Q$ factor are important for laser technology, we investigated the limiting case with and the refractive index of the defect layer $n^d = \sqrt{(\varepsilon_1 + \varepsilon_2)/2}$. Figure 4 shows the dependence of ln$Q$ on the CLC layer thickness ($d/\sigma$). The defect is in the midplane of the system. One can see that this dependence is linear; correspondingly, high ln$Q$ factors can be obtained by increasing the CLC thickness.

## 3. CONCLUSIONS

We showed that the dependence of ln$Q$ on the CLC layer thickness is linear and that a change in the refractive index of the medium adjacent to the CLC layer significantly changes the reflection at the defect mode and the defect-mode wavelength. The results obtained can be used to design miniature low-threshold lasers with a circularly polarized fundamental mode, narrow-band filters and mirrors, optical diodes, etc.

_